# Human-Human interaction: epidemiology

Andrzej Jarynowski[a,b]

[a]Smoluchowski Institute of Physics, Jagiellonian University,
30-059 Kraków, ul. Reymonta 4, Poland
[b]UNESCO Chair of Interdisciplinary Studies, Wrocław University,
50-204 Wrocław, pl. M. Borna 9, Poland

Abstract

The aim of this work was to show few examples and few perspective of modeling in epidemiology. We began with differential equations which were a first tool to describe and predict that phenomena. Wroclaw as a cite was very important, because statistics from smallpox epidemic were used by Bernoulli to estimate parameters of first mathematical model of epidemic. Next step were SIR models and those also appeared first as differential equations. They were very popular in begin of XX century. When computer simulation changed the world of mathematical modeling agent-based models gave more possibilities in epidemiology. That models have a big privilege on differential equation, because of information of social network people habits and reaction on infections, which can me involved in agent-based models as well as governmental intervention. We showed in this work how that human relations are important in transmitting diseases and there is example, where it is possible to conduct experiments of significant policy relevance (vaccinating), such as investigating the initial growth of an epidemic on a real-world network. Presented H1N1 model cauld be observed in real time (prediction was made in September 2009 for winter season 2009/2010) what make it more exiting and also practice.

## 4.1) Introduction

This area is explored by researchers from different academic background: physicians, physicists, mathematicians, statisticians, computer scientists and sociologists. All of them have something to add and this work based on different base in different parts from mathematical beginning though computer science (simulations) and concept from physics to finish with sophisticated sociological and statistical analyses. Aim of our work was first to collect few perspective (mostly mathematical) and developed them, but in effect final analyze is more empirical, what make it more practical in use.  This work is also very specious, because of its interdisciplinary and showing light on methods which were not use in epidemiology till this moment. Epidemiological models that treat transmission as "human-to-human" from differential equation point of view do exist in older literature, but in more recent agent-based models appear more often[1]. We have started from general idea of models in this area with fascinating role of our city-Wrocław, then turn to differential equation models and finish with agent-based models. Mathematical models and computer simulation start to play significant role as quantity of social interactions is enormous, but more important than simulations are real data especially register-based. Simulation has sense only if there are some data you can calibrate parameters to run yours simulations. This need enables cooperation between registering institutions which exerts a pressure on collecting data for simple analyse with many researchers who work on new models and use complex tools taken often from other disciplines. Our agent-based models were calibrated on Swedish data, because that country has a huge dataset of citizen's activities and give opportunity to analyze them. In both cases we show that one year is a typical time window as it was showed in prevous article of this book.

## 4.2) Epidemics

An *epidemic* (from Greek epi- upon + demos people) an occurrence of disease that is temporarily of high prevalence. The rise and decline in epidemic prevalence of an infectious disease is a probability phenomenon dependent upon transfer of an effective dose of the infectious agent from an infected individual to a susceptible one[1]. The same phenomena is called *epizootic* by animals and *epizitofic* by plants. In recent usages, the disease is not required to be communicable; examples include cancer or heart disease, but we concentrate only those with transmission in this work. If epidemic occurring over a wide geographical area (e.g, worldwide) is called a *pandemic*. Nowadays we have a lot of examples of pandemics and this problem will be solving. Who didn't hear about swine flu during spring 2009 or bird flu in 2005?

Science, which cope with epidemics is epidemiology. Epidemiological studies may be classified as descriptive or analytic. In descriptive epidemiology, demographic surveys are used to determine the nature of the population affected by the disorder in question, noting factors such as age, sex, ethnic group, and occupation among those afflicted. Other descriptive studies may follow the occurrence of a disease over several years to determine changes or variations in incidence or mortality; geographic variations may also be noted. Descriptive studies also help to identify new disease syndromes or suggest previously unrecognized associations between risk factors and disease. In addition to providing clues to the causes of various diseases, epidemiological studies are used to plan new health services, determining the incidence of various illnesses in the population to be served, and to evaluate the overall health status of a given population. In most countries of the world, public-health authorities regularly gather epidemiological data on specific diseases and mortality rates in their populaces[2]. Main field of their studies focus on non-interacting systems. Most of knowledge came from analytical analysis and tests. These studies divide a sample population into two or more groups, selected on the basis of suspected causal factor (for example, cigarette smoking) and then monitor differences in incidence, mortality, or other variables. This way of studying statistical properties can be supplemented by modeling. Mathematical models can help in describing phenomena of epidemiology spreading and give an answer how to fight with them. There are many problems in modeling because of big variety of epidemic types. Many of epidemics appeared in developed countries, such as colds or influenza, cause only temporal disease to most sufferers, although economically (in terms of medical cost or employment) their effects may be substantial. On the other hand in undeveloped countries experienced diseases such cholera, and cause large numbers of death, especially if treatment is not available.

---

1  Encyclopaedia Britannica *"Epidemic"*

### 4.3) Historical aside on epidemics

During ages appear phenomena, which is called by demographers epidemiological transition. The main process is a transition from communicable or infectious diseases to man-made disease as a cause of mortality. In deed, knowledge of causes of death in less developed population based on bacteria and virus. Data provided by UN in typical population with big proportion of young people and quit big mortality about 35% of death are caused by infectious and respiratory diseases, 25% cardiovascular diseases and cancer. Instead that in typical population with big proportion of old people and quit low mortality with first group of death is connected less than 5% of deaths, with second more than 80%. According to this concept dynamic of population change in not cause by natality, but by morbidity. The same oscillation are observed in animal world. Mathematical model of Volterra-Lotka shows how 2 populations of predator and pray are changing in time. That extraordinary deaths are sometimes called catastrophic and in human world ware caused by epidemics, famine etc. A. R. Orman suggested sequence of 3 types of mortality[2] : age of plagues (big fluctuations in cycles of famine and epidemics); age of expiring epidemics (much more systematic deaths and much less fluctuations); age of man-made disease (stable level of deaths). We can think, then in last phase epidemics should not appear. Unfortunately old people in modern societies are more vulnerable to infectious diseases. They often live in crowded cities and infectious diseases can get in to synergy with old-people-diseases. Parallel appeared new species resistant to drugs. They caused growth in some almost forgotten diseases. For example Ebola virus was discovered in $70^{th}$; Human immunodeficiency virus (HIV) and Hepatitis C in $80^{th}$; hemorrhagic fever in $90^{th}$. Their role in mortality is quit big, with notation that AIDS is $4^{th}$ mainly cause of death on earth.

---

2  Encyclopaedia Britannica *"Epidemiology"*

## 4.4) Bernoulli's model [3]

> *I simply wish that, in a matter which so closely concerns the well-being of mankind, no decision shall be made without all the knowledge which a little analysis and calculation can provide.*
> Daniel Bernoulli, presenting his estimates of smallpox (Royal Academy of Sciences-Paris, 30 April 1760)

This Swiss mathematician was the first to express the proportion of susceptible individuals of an endemic infection in terms of the force of infection and life expectancy. His work describe smallpox, which cause a lot of epidemics in big European cities at his time. Smallpox devastated earlier the native Amerindian population and was an important factor in the conquest of the Aztecs and the Incas by the Spaniards. In Poland smallpox last time appear in Wroclaw in 1963, but it was stopped by the actions of the government and epidemiologists. Moreover smallpox was eradicated by WHO in 1979. Bernoulli actually used date provided from Wroclaw to estimate his model. He based at work of famous British astronomer – Edmund Halley[4]. In $17^{th}$ century English Breslaw means Breslau-German name of Wroclaw. In beginning of his publication Halley wrote his purpose: „This *Defect* seems in a great measure to be satisfied by the late curious Tables of the Bills of *Mortality* at the City of *Breslaw,* lately communicated to this Honorable Society by Mr. *Just ell*, wherein both the *Ages* and *Sexes* of all that die are monthly delivered, and compared with the number of the *Births*, for Five Years last past 1687, 88, 89, 90, 91, seeming to be done with all the Exactness and Sincerity possible." Later he described Wroclaw: „This City of *Breslaw* is the Capital City of the Province of *Silesia*; or, as the *Germans* call it, *Schlesia*, and is situated on the Western Bank of the River *Oder*, anciently called *Viadrus*; near the Confines of *Germany* and *Poland (...)* It is very far from the Sea, and as much a *Mediterranean* Place as can be desired, whence the confluence of Strangers is but small, and the Manufacture of Linnen employs chiefly the poor People of the place, as well as of the Country round about; whence comes that sort of Linnen we usually call your Schlesie *Linnen*; which is the chief, if not the only Merchandize of the place. For these Reasons the People of this City seem most proper for a *Standard*; and the rather, for that the *Births* do, a smaller matter, exceed the *Funerals*. The only thing wanting is the Number of the whole People, which in some measure I have endeavored to supply by comparison of the *Mortality* of the People of all Ages, which I shall from the said Bill traces out with all the Accuracy possible."

Bernoulli's main objective was to calculate the adjusted life table if smallpox were to be eliminated as a cause of death. His formula is valid for arbitrary age-dependent host mortality, in contrast to some current formulas which underestimate herd immunity.

Bernoulli wrote:

*(1) I shall assume that, independent of age, in a large number of persons who have not had smallpox, the disease annually attacks one person out of as many persons as there are units in the number n. According to this hypothesis, the danger of catching the disease would remain the same at every age of one's life, provided that one had not already had it. If, for example, we put n=10, it would be the fate of every person to be decimated every year of his life in order to know whether he would have smallpox this year or not, right up to the moment when that fate actually befell him. This hypothesis seems to me to be very probable for all young people up to the age of sixteen to twenty years. If we see few people over that age who catch smallpox, it is because the great majority have already been infected by it. What follows will enable us to see what degree of probability this hypothesis merits.*

*(2) In the second place, I shall assume that, at whatever age one catches smallpox, the danger of dying of it is always the same and that out of a number of patients expressed by the number m, one dies of it. With regard to this assumption, I note that no doctor would take it into his head to suppose that, other things being equal, smallpox is more or less dangerous merely on account of the age at which it is caught, provided that this age does not exceed twenty. It is only above this age that we usually suppose that smallpox becomes a little more dangerous. We shall later have occasion to examine this hypothesis more closely.*

Let set some variable to describe his model:
− the present age, expressed in years=$x$;
− the number of survivors at this age=$\zeta$;
− the number of those who have not had smallpox at this age=$s$;
− and let us retain the meaning given above to the letters $m$ and $n$.

Here is the reasoning which can be followed to find a general expression for.

$$-ds = \frac{sdx}{n} - \frac{sd\zeta}{\zeta} - \frac{ssdx}{mn\zeta}$$  (4.1)

Change of $s$ is negative, because $s$ is decreasing in time. On the right hand side we have: number of people who catch smallpox in time interval; number of people who died.

We can derive that equation doing substitution:

If we put $\frac{\zeta}{s} = q$, (4.2)

we have (4.3)

$$\frac{dq}{dx} = \frac{q}{n} - \frac{s}{mn}$$

and hence finally $\quad s = \dfrac{m}{e^{\frac{x+C}{n}}+1}\zeta \quad$ (4.4)

Putting initial condition at $x=0$, $s=\zeta$ (each letter then expressing the number of newborn children involved)

$$s = \dfrac{m}{(m-1)e^{\frac{x}{n}}+1}\zeta \quad (4.5)$$

Bernoulli put in his consideration $m=8$ and $n=8$.

$$s = \dfrac{8}{7e^{\frac{x}{8}}+1}\zeta \quad (4.6)$$

| Age in years | Survivors according to Halley | Not having had smallpox | Having had smallpox | Catching smallpox each year | smallpox each year | Totall smallox deaths | Death of other diseases each year |
|---|---|---|---|---|---|---|---|
| 0 | 1300 | 1300 | 0 | | | | |
| 1 | 1000 | 896 | 104 | 137 | 17,1 | 17,1 | 283 |
| 2 | 855 | 685 | 170 | 99 | 12,4 | 29,5 | 133 |
| 3 | 798 | 571 | 227 | 78 | 9,7 | 39,2 | 47 |
| 4 | 760 | 485 | 275 | 66 | 8,3 | 47,5 | 30 |
| 5 | 732 | 416 | 316 | 56 | 7 | 54,5 | 21 |
| 6 | 710 | 359 | 351 | 48 | 6 | 60,5 | 16 |
| 7 | 692 | 311 | 381 | 42 | 5,2 | 65,7 | 12,8 |
| 8 | 680 | 272 | 408 | 36 | 4,5 | 70,2 | 7,5 |
| 9 | 670 | 237 | 433 | 32 | 4 | 74,2 | 6 |
| 10 | 661 | 208 | 453 | 28 | 3,5 | 77,7 | 5,5 |
| 11 | 653 | 182 | 471 | 24,4 | 3 | 80,7 | 5 |
| 12 | 646 | 160 | 486 | 21,4 | 2,7 | 83,4 | 4,3 |
| 13 | 640 | 140 | 500 | 18,7 | 2,3 | 85,7 | 3,7 |
| 14 | 634 | 123 | 511 | 16,6 | 2,1 | 87,8 | 3,9 |
| 15 | 628 | 108 | 520 | 14,4 | 1,8 | 89,6 | 4,2 |
| 16 | 622 | 94 | 528 | 12,6 | 1,6 | 91,2 | 4,4 |
| 17 | 616 | 83 | 533 | 11 | 1,4 | 92,6 | 4,6 |
| 18 | 610 | 72 | 538 | 9,7 | 1,2 | 93,8 | 4,8 |
| 19 | 604 | 63 | 541 | 8,4 | 1 | 94,8 | 5 |
| 20 | 598 | 56 | 542 | 7,4 | 0,9 | 95,7 | 5,1 |
| 21 | 592 | 48,5 | 543 | 6,5 | 0,8 | 96,5 | 5,2 |
| 22 | 586 | 42,5 | 543 | 5,6 | 0,7 | 97,2 | 5,3 |
| 23 | 579 | 37 | 542 | 5 | 0,6 | 97,8 | 6,4 |
| 24 | 572 | 32,4 | 540 | 4,4 | 0,5 | 98,3 | 6,5 |

**Table 4.1.** Smallpox in Wroclaw (based on Bernoulli [3])

Bernoulli has constructed the [Tab. 4.2] at the end of this Memoir, in which the first two columns are the same as in the [Tab. 4.1], though he has given the second column another name, 'natural state with smallpox', in contradistinction to the third column, which shows the 'state without smallpox' and which gives the number of survivors each year assuming that nobody must die of smallpox. Difference between second and third column gave him a gain in people's lives. He introduced 'total quantity of life' of the whole generation, for each of the two states, for the sum of all the numbers of the second column and of the third column respectively [Tab. 4.2]. Making this deduction, we obtain the total quantity of life for the state free from smallpox, with the whole tribute paid, which must be compared with tribute for the natural state. If take expected values of variables in second in third column (sum all multiplication of age and values in column and we divide these numbers by 1300) we will have the average life for the natural state as 26 years 7 months, for the state without smallpox and without tribute as 29 years 9 months and for the state free from smallpox. Under these assumptions an individual's expectation of life at birth would increase from 26 years 7 months to 29 years 9 months.

| Age in years | Natural state with smallpox | State without smallpox |
|---|---|---|
| 0 | 1300 | 1300 |
| 1 | 1200 | 1171 |
| 2 | 855 | 881,8 |
| 3 | 798 | 833,3 |
| 4 | 760 | 802 |
| 5 | 732 | 779,8 |
| 6 | 710 | 762,8 |
| 7 | 692 | 749,1 |
| 8 | 680 | 740,9 |
| 9 | 670 | 734,4 |
| 10 | 661 | 728,4 |
| 11 | 653 | 722,9 |
| 12 | 646 | 718,2 |
| 13 | 640 | 741,1 |
| 14 | 634 | 709,7 |
| 15 | 628 | 705 |
| 16 | 622 | 700,1 |
| 17 | 616 | 695 |
| 18 | 610 | 689,6 |
| 19 | 604 | 684 |
| 20 | 598 | 678,2 |
| 21 | 592 | 672,3 |
| 22 | 586 | 666,3 |
| 23 | 579 | 659 |
| 24 | 572 | 651,7 |
| 25 | 565 | 644,3 |

**Table 4.2.** Gain in life [Smallpox in Wroclaw (based on Bernoulli [3])

There are some parameters, which are used today, introduced by Bernoulli: force of infection $1/n$ (the annual rate of acquiring an infection) the case fatality $1/m$ (the proportion of infections resulting in death). From several large cities (not only Wroclaw) which recorded cause-specific numbers of deaths, estimates of $1/m$ were known to be about 1/13 so 7.7%. Bernoulli used Halley's life table for the city of Wroclaw and came up with the estimates $1/n =1/8$ and also $1/m=1/8$ so 12.5%. For Paris he assumed a life expectancy of 32 years which yields a proportion of susceptible individuals of 15%.

### 4.5) Differential equations of epidemics

Mathematical description of phenomena needs more assumptions has to be made in terms of understanding real situation. In first instance let consider the spread of a non-fatal disease, to which no-one is naturally immune. Suppose the population can be divided into two groups: Susceptible-Healthy and Infectious-Infected [5].
Assume that at general time *t*:
$S(t)$ = Number of Susceptible
$I(t)$ = Number of Invectives
with $S(t)+I(t) = N$
The problem now to model spread of the disease.
Consider a single *susceptible* individual in a homogeneously mixing population. This individual contacts other members of the population at the rate $C$ (with units time$^{-1}$) and a proportion $I/N$ of these contacts are with individuals who are infectious. If the probability of transmission of infection given contact is $\beta$, then the rate at which the infection is transmitted to *susceptibles* is $\beta CI/N$, and the rate at which the *susceptible* population becomes *infected* is $\beta CSI/N$.

The *contact rate* is often a function of population density, reflecting the fact that contacts take time and saturation occurs. One can envisage situations where $C$ could be approximately proportional to $N$ (which corresponds to mass action), and other situations where $C$ may be approximately constant. Hence terms like $\beta SI$ and $\beta SI/N$ are frequently seen in the literature. For these, and in many instances where the population density is constant, the contact rate function $C$ has been subsumed into $\beta$, which is now no longer a probability but a "transmission coefficient" with units time$^{-1}$. To reduce number of coefficient let write: $r = \beta C/N$. The number of invectives at time *t* is given by the differential equation:

$$\frac{dI}{dt} = rSI \qquad (4.7)$$

That simple model show, that epidemics spread rapidly through a closed community, when it is reasonable to assume that there are no recoveries during the period of epidemic. However it is not valid in most cases. Thus development of model allow possibility of recovery. Following recovery, and individual either becomes immune to the disease or again becomes susceptible to it.

Initially we will couple with second case. Looking at the recovery term we assume that it is proportional relationship with invectives. It gives:

$$\frac{dI}{dt} = rSI - aI \qquad (4.8)$$

In next step suppose that population is divided into three classes: the susceptibles (*S*) who can catch the disease; the infectives (*I*), who can transmit disease and have it; and the removed (*R*) who had the disease and are recovered (with immune) or isolated from society. Schema of transition can be represented[6]:

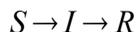

$$S \rightarrow I \rightarrow R$$

The model mechanism may be updated to following form:

$$\frac{dI}{dt} = rSI - aI, \quad (4.9)$$

$$\frac{dS}{dt} = -rSI, \quad (4.10)$$

$$\frac{dR}{dt} = aI \quad (4.11)$$

A key question in for given $r$, $a$, $S_0$, $I_0$, whether the infection will spread or not and if it how it develops in time and when it start to decline. Since initial condition on $S$- $S_0 < a/r$ then $dI/dt < 0$ in which case $I_0 > I(t)$ and $I$ goes to 0 with $t$ going to infinity. On the other hand if $S_0 > a/r$ then $I(t)$ increase and appear epidemic. We have something like threshold phenomenon $S_c$ and depend on level of initial numbers. Inversion of this critical parameter is infection's *contact rate* $(=r/a)$.
We can write:

$$R_0 = \frac{r S_0}{a} \quad (4.12)$$

where $R_0$ is basic reproduction rate of the infection. This rate is crucial for dealing with and an epidemic which can be under control with vaccination for example. Action is needed if $R_0 > 1$, because epidemic clearly ensues then. We have be know

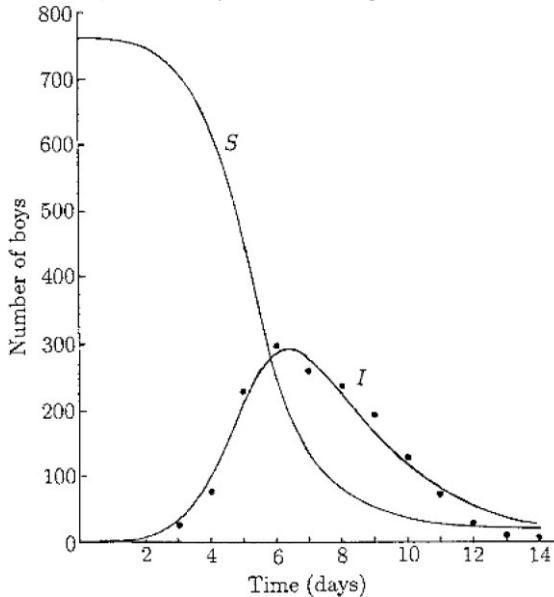

**Figure 4.1.** Influenza epidemic data (●) for a boys boarding school as reported in UK in 1978. The continous curves for $I$ and $S$ were obtained from a best fit numerical solution of SIR system with parameters: $S_0 = 762$, $I_0 = 1$, $S_c = 202$,

$r=0,0022$/day. The condition of epidemic occur, $S_0>S_c(=a/r)$ [graph by J. Murray[3]]

Other assumption for developing model is to incorporate the possibility of replenishment in the number of healthy susceptibles.

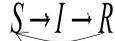

This may be in form of new births, immigration or by the loss of immunity of some whose have had the disease and recovered. On the basics of a constant replenishment rate $u$ which is new in model ($u$ is only difference between 4.10, 4.11 and 4.14, 4.15).

$$\frac{dI}{dt}=rSI-aI,\quad (4.13)$$

$$\frac{dS}{dt}=-rSI+u,\quad (4.14)$$

$$\frac{dR}{dt}=aI-u \quad (4.15)$$

It is rapidly seen that $S(t)=a/r$ and $I(t)=u/a$ are an equilibrium state corresponding to 4.13-15.
Of course this model can be more complicated, but without change in this system of equations. Our parameter $u$, which describes replenishment from $R$ to $S$ can be understand as a migration rate and taking initial condition of $R$ we have possibility to build reservoir of population, which can migrate into system.
That few simple models are only beginnings of this work and give one base for further consideration and more sophisticated (but not in terms of differential equation) models.
$R_0$ as the basic reproduction rate must be established on time interval. Dynamic properties of epidemic bring problems with validity of proposed $R_0$. It has to be optimized (cannot obey only short time scale or only long). Concepts from introduction of whole book can be used for that.

## 4.6) Consequences to Society of H1N1 Pandemic in Sweden[4]

Traditionally SIR-models have been used to aid the understanding of these processes (what we were showing in previous chapters). SIR is a compartmental model in which differential equations govern the dynamic flow between three compartments and no contact structure is assumed. In an SIR-type model, the population is split into three different groups and the majority of the population is placed in the susceptible compartment. All information about society is used in microsimulation, so it can give better prediction, then differential equations. Simplified compartmental models provide inadequate representations because contacts between susceptible and infectious persons are not random[8]. We can use a lot of information in this consideration, which are provided by government. Another reason is that, there is possible to run algorithms for society of whole country-Sweden and there are not too time consuming, e.g. one 180 day simulation take about less than 1h on personal computer.

---

3  Reference [6]. You can find more historical example, where epidemic statistics fit to model as here. This one looks like ideal epidemic curve.
4  I was involved in that project. Results were used in governmental purpose and published in Swedish [7].

Database of programme takes account data from:

1. National population register- all information about sex, age, etc.
2. Employment register- it provides list of employee or students
3. Geography database -100x100 coordinates of houses, schools, hospitals or workplaces

Disease transmission is performed twice daily at 9 am and 5 pm. Programme checks where are all persons during that day hours and night hours. The house member lists and patient lists are iterated to calculate the combined infectiousness of their members. In the case of larger workplaces, the member lists are further divided into departments. After the combined infectiousness is determined, the list members are exposed and infected according to infection risk.

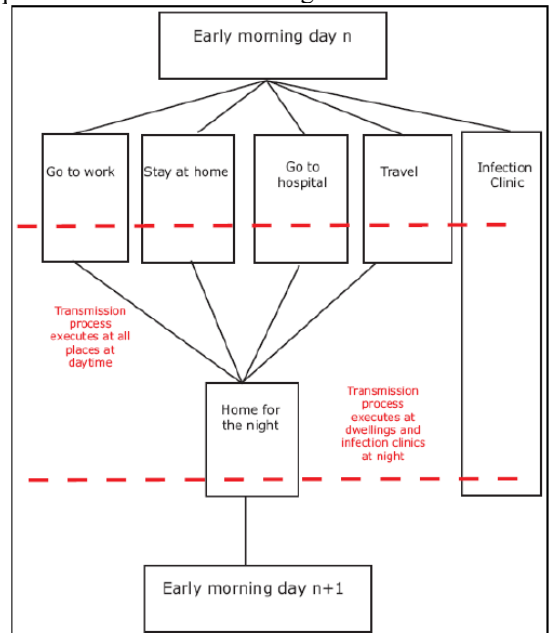

**Picture 1**. The daily routines for the simulated persons. [pic. by Brouwers][5]

Additional contacts (and transmission) are included in model in two ways (Sweden

---

5   Brouwers [9]

is divided into 81 regions with their own characteristics of density or intensivity of traveling):
1. Neighbourhood
2. Travel[6]

Infectionous scheme depends on level of illness. Carrat[11] introduced profile which was used by authors of model in sense of Weibull distribution. After some calibration it was found that the best probabilities give transmission coefficient $R_0$ at level 1,4.

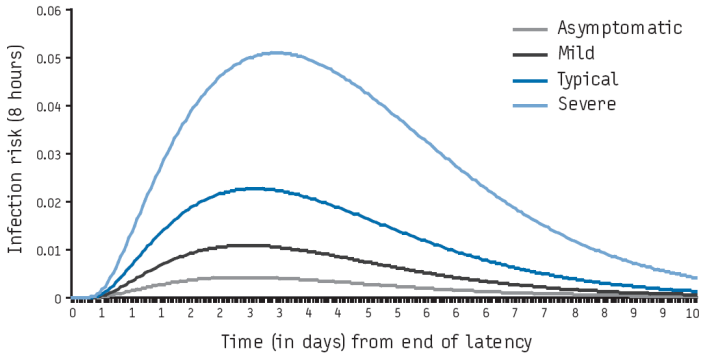

**Figure 4.2.** Profile of probability of sending infections for adults in MicroSim model [adopted by Brouwers[7]]

The simulations were run with the following assumptions. The outbreak of pandemic influenza in Sweden starts depend of method in June or in September. $R_0$-value corresponding approximately to 1,4 in main model, because that was observed in New Zealand during their outbreak. The viral infectivity is markedly initially $R_0$ value of approximately 2,1 in preliminary method.

Formula for $R_0$ given in chapter about differential equations is redefined as the average number of individuals a typical person infects under his/her full infectious.

$$R_0 = \frac{-\ln(\frac{A}{S_0})}{1 - \frac{A}{S_0}} \quad (4.16)$$

Immunity was calibrated in model to obtain $R_0 = 1,4$
$S_0$: Total number of susceptible individuals before the outbreak
$A$: Total number of susceptible individuals after the outbreak
Formula (4.16) is defined as the average number of individuals a typical person infects under his/her full infectious and its aproximation of reproduction rate from previous chapter (4.12). Let remain at this point the goal of whole book. $R_0$ is oriented on time interval. Difficulty with establishment of this value comes from seasonality of influenza. The main Swedish epidemiologist established $R_0$ from previous season in calendar winter of New Zealand due to similarity of this country in terms of size (population and geographic) or special aspect of isolation.

---
6 Swedish transport data was from governmental report [10]
7 Brouwers [12] One of infectivity profiles (there are more in original paper)

To compare the societal costs of the scenarios, the following costs—obtained from health economists at the Swedish Institute for Infectious Disease Control (SMI)—were used:

- Cost of one day's absence from work, for a worker: SEK 900.
- Cost of treatment by a doctor in primary care: SEK 2000.
- Cost of one day's inpatient care: SEK 8000.
- Cost of vaccine and administration of vaccination for one person: SEK 300.

For all scenarios, the SEK 300 vaccine costs are based on the assumption that the entire population is vaccinated (a total of 18 million doses), split evenly between vaccine cost and vaccine administration. This means that no savings on vaccine administration can be attributed to a lower number of vaccinated than 90%. The model presupposes absent workers to take care of sick children, and thus sick children do not produce the SEK 900 cost in a family where a parent is already ill. The following scenarios were compared: no response, the vaccination coverage of 30%, 50%, 60%, 70% or 90%. Each simulation take 180 days and starts with 50 randomly selected infected individuals at day 0. Day 0 in the model corresponds to the first September 2009. Some scenarios were investigated during 300 days, this was where it was interesting to see whether the outbreak was going to appear after 180 days.

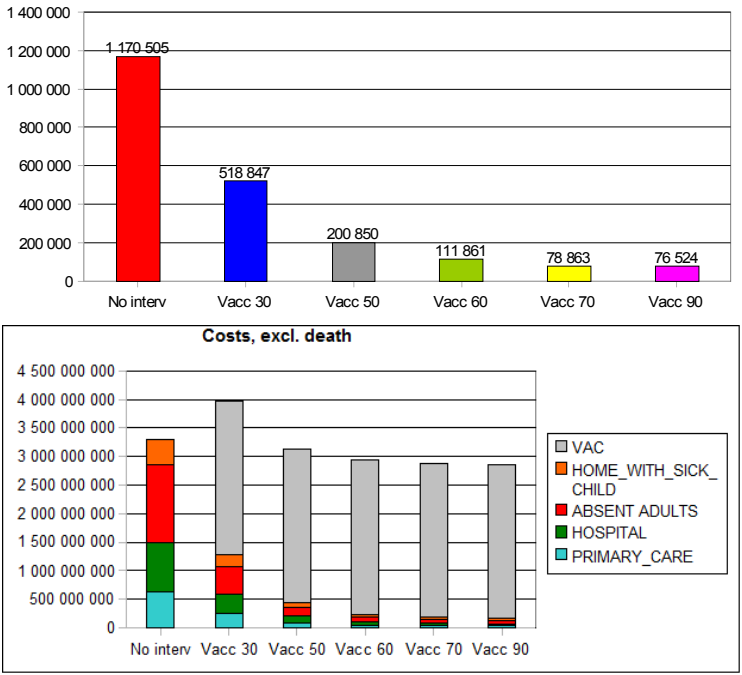

**Figure 4.3.** No of infections (upper) and total cost of influenza excluding cost of death cases in SEK (lower) [graphs by Brouwers[8]]

This model is one of the first[13], which are using individual social properties of society and it is first one, which works on scale of whole country. We have to careful with interpret the results. It was announced to the media and if You look at official report, You will find that it was censored to avoid public panic. One thing is very important-Swedish politics makers have seen it and they gave feedback, which make this model more realistic. One of conclusion of the report was to vaccinate more than 70% of population. In Sweden, vaccination is voluntary, but for the purpose of these simulation experiments it was assumed, somewhat unrealistically. It is difficult to say when appear really outbreak, but when we look at real data we can see, that model is almost correct to this time with following suggestions: Authors hypothesis is that the relatively rapid, especially in view of the $R_0$ values reported, peaks in Australia and New Zealand could be explained by the earliest cases going unrecognized, and a constant influx of new cases from abroad.

---

8   Brouwers [7] English translation of graph description

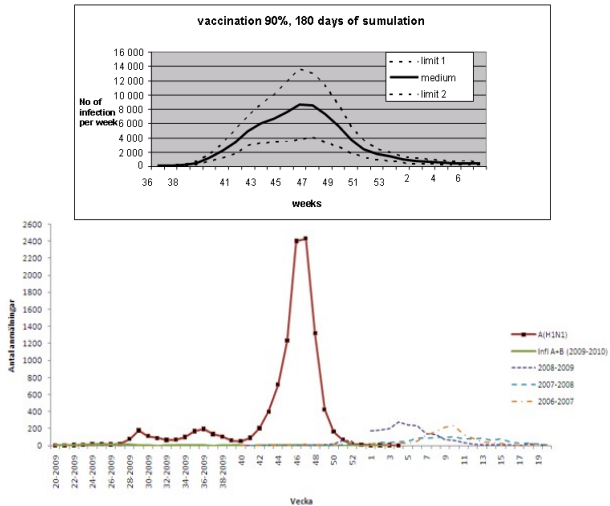

**Figure 4.4.** No. of infections per week in simulation (upper graph by L. Brouwers[9]) and cases collected by The Swedish Institute for Infectious Disease Control (lower graph by SMI[10]).

## 4.7) Summary

The aim of this work was to show few examples and few perspective of mathematical modeling in epidemiology. We began with differential equations which were a first tool to describe and predict that phenomena. Wroclaw as a city was very important, because statistics from smallpox epidemic were used by Bernoulli to estimate parameters of first mathematical model of epidemic. Next step were SIR models and those also appeared first as differential equations. They were very popular in begin of XX century. When computer simulation changed the world of mathematical modeling agent-based models gave more possibilities in epidemiology. That model have a big privilege on differential equation, because of information of social network, people habits and reaction on infections, which can me involved in agent-based models as well as governmental intervention. We showed in this work how that human relation are important in transmitting diseases and there are institutions, whose aim are to decrease cost of epidemic in societies (like ECDC: European Centre for Disease Prevention and Control). In recent years, the knowledge of social networks experienced an accelerating growth. To solve more complex and sophisticated social problems, new types of tools and models are constantly developed. Simulation modeling is very important because it provides the insight into the dynamics of the social process whereas commonly used methods of research not always make it possible. Moreover, such models allow not only to test theory but also they are inspiring and may support constructing new theories in sociology. The most interesting is the possibility to study the dynamics of collective behaviour itself and the relationships among variables of interest.

---

9     Brouwers [7] English translation of graph description
10    SMI: Numbers of positive test on H1N1 of infections are reported weekly on website http://www.smi.se. There is opinion, that only 1/10 of all cases of infection is tested positive in labs. In that terms simulation gives the same result as real life.

As a example we presented H1N1 pandemic in winter 2009/2010. This very sophisticated model is first in history, which calculate states of all agents which represent all citizen of Country (Population of Sweden is about 9 millions). We could observe last winter how this model is suitable to real situation. This is the best prove of goodness of that model. That shows, that mathematical modeling can by useful and gives answer (in this case about vaccination of population).